\documentclass[final]{svjour3}
\usepackage{graphicx}
\usepackage{bm}
\usepackage{amssymb}
\usepackage{mathptmx}
\usepackage[numbers]{natbib}
\usepackage{float}
\usepackage[hypertex]{hyperref}

\newcommand{\qr}{{\bf r}}

\def\bra#1{\bigl\langle{ #1} \bigr|}
\def\ket#1{\bigl|{ #1} \bigr\rangle}
\def\Bra#1{\Bigl\langle{ #1} \Bigr|}
\def\Ket#1{\Bigl|{ #1} \Bigr\rangle}
\def\ovlp#1#2{\bigl\langle{ #1}\big|{#2} \bigr\rangle}

\def\half{\frac{1}{2}}
\def\rvec{{\bf r}}
\def\evec{{\bf e}}

\def\kvec{{\bf k}}
\def\pvec{{\bf p}}
\def\qvec{{\bf q}}

\def\I{\mathrm{i}}

\def\he#1{$^{#1}$He}
\def\I{\mathrm{i}}
\journalname{Journal of Low Temperature Physics}

\begin{document}
\title{Transport and Phonon Damping in $^{\bf 4}$He }
\author{K. Beauvois$^{\dagger\ddagger}$, H. Godfrin$^\ddagger$, E.\ Krotscheck$^{+\#}$
 R. E. Zillich$^\#$}
\institute{$^{\dagger}$Institut Laue-Langevin, 6, rue Jules Horowitz, 38042 Grenoble, France\\
$^{\ddagger}$Univ. Grenoble Alpes, CNRS, Grenoble INP, Institut N\'eel, 38000 Grenoble, France\\
$^+$Department of Physics, University at Buffalo SUNY,
Buffalo, NY 14260, USA\\
$^\#$Institut f\"ur Theoretische Physik, Johannes
Kepler Universit\"at, A 4040 Linz, Austria}

\maketitle
%
%
%
\begin{abstract}
  The dynamic structure function $S(k,\omega)$ informs about the
  dispersion and damping of excitations. We have recently
  (Phys. Rev. B {\bf 97}, 184520 (2018)) compared experimental results
  for $S(k,\omega)$ from high-precision neutron scattering experiments
  and theoretical results using the ``dynamic many-body theory''
  (DMBT), showing excellent agreement over the whole experimentally
  accessible pressure regime. This paper focuses on the specific
  aspect of the propagation of low-energy phonons. We report
  calculations of the phonon mean-free path and phonon life time in
  liquid \he4 as a function of wave length and pressure.
  Historically, the question was of interest for experiments of
  quantum evaporation. More recently, there is interest in the
  potential use of \he4 as a detector for low-energy dark matter
  (K. Schulz and Kathryn M. Zurek, Phys. Rev. Lett. {\bf 117}, 121302
  (2016)).  While the mean free path of long wave length phonons is
  large, phonons of intermediate energy can have a short mean free
  path of the order of $\mu$m. Comparison of different levels of
  theory indicate that reliable predictions of the phonon mean free
  path can be made only by using the most advanced many--body method
  available, namely, DMBT.

\end{abstract}

\section{Introduction}
\label{sec:introduction}

It has been known for a long time
\cite{MarisRMP}\nocite{PhysRevA.8.1980,Rugar1984}-\cite{Sridhar87}
that low-energy phonons in liquid \he4 display an anomalous dispersion
relation which allows these phonons to decay.  Precise neutron
scattering measurements of the phonon dispersion in liquid $^4$He
\cite{skwpress} provide the phonon dispersion relation for all
experimentally accessible densites with unprecedented accuracy. They
confirm the finding of earlier work that the phonon dispersion
relation is anomalous up to densities of about $\rho =
0.0245$\AA$^{-3}$. These data agree very well with our recent theoretical
results for the dynamics of $^4$He based on time-dependent
multiparticle correlations \cite{eomIII}; our methods should therefore
also be capable of quantitative microscopic predictions for the phonon
lifetime.  We apply our many-body theory of inelastic scattering,
previously applied to scattering off $^4$He droplets \cite{dropdyn}
and the surface of $^4$He
\cite{hscatt}. In the following sections, we briefly review both the
theoretical methods used to calculate the relevant ground state
properties (Section \ref{sec:groundstate}) and the dynamic features
(Section \ref{sec:DMBT}) of the \he4 liquid, and the data analysis
to obtain accurate values of the phonon dispersion coefficient.

The investigations are, among others, of interest because multiple
scattering in superfluid $^4$He has been proposed as a detection
mechanism for low-mass dark matter
\cite{PRL117Zurek}\nocite{PRD95Zurek,MarisDarkMatter}-\cite{Romani2019},
among other proposed detectors \cite{PhysRevLett.119.131802}.  Such a
detector design requires accurate knowledge of the propagation of
low-energy phonons within the medium. In particular, when the phonon
dispersion relation is anomalous, phonons are damped and have a finite
mean free path.

\section{Ground State Structure of $^4$He}
\label{sec:groundstate}
Microscopic calculation of properties of many--body systems
begin with an accurate calculation of ground--state properties.
For \he4 it is adequate to begin with a non-relativistic
Hamiltonian
\begin{equation}
  H = -\frac{\hbar^2}{2 m}\sum_i \nabla_i^2 + \sum_{i<j} v(|\qr_{i}-\qr_j|)
\end{equation}
where the pair-wise interaction $v(r)$ is taken from Aziz {\em et al.\/}
\cite{AzizII}.

\begin{figure}

\centerline{\includegraphics[width=0.45\textwidth,angle=-90]{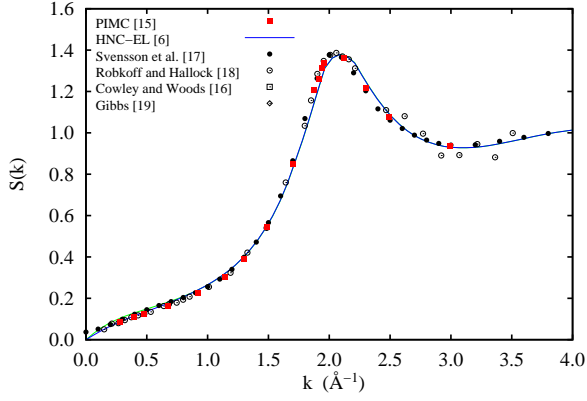}}
\caption{(color online) Static structure function $S(k)$ for $^4$He at
  equilibrium density. We compare JF-EL, Monte Carlo
  \cite{JordiQFSBook}, and experiments
  \cite{CowleyWoods,svensson,RoHa,AndersenRoton}. The figure is from
  Ref. \citenum{eomIII}.  }
\label{fig:scompare}
\end{figure}

The most efficient evaluation of ground state properties
is done by the variational Jastrow-Feenberg ansatz for the ground state:
\begin{equation}
\Psi_0({\bf r}_1,\ldots,{\bf r}_N) =
 \exp\,\frac{1}{2}
        \left[\sum_{i<j} u_2({\bf r}_i,{\bf r}_j) + \sum_{i<j<k} u_3({\bf r}_i,{\bf r}_j,{\bf r}_k)\right]\,.\label{eq:wavefunction}
\end{equation}
The correlation functions $u_i(\qr_1,\dots,\qr_i)$ are obtained by
minimizing the ground state energy $E_0$
\begin{equation}
\frac{\delta}{\delta u_i(\qr_1,\dots,\qr_i)}
\frac{\bra{\Psi_0}H\ket{\Psi_0}}{\ovlp{\Psi_0}{\Psi_0}} = 0\,.
\end{equation}
The method is known as Jastrow-Feenberg-Euler-Lagrange (JF-EL) method.
The key quantity obtained from such a ground state calculation and
used for the calculation of the dynamics is the static structure
function $S(k)$. We show in Fig. \ref{fig:scompare} a comparison of
different calculations and experiments.

\section{Many-Body Dynamics}
\label{sec:DMBT}

Dynamics is treated at the same level as the ground state: We write
the dynamic wave function as containing a small, time-dependent
component
\begin{equation}
        \left|\Psi(t)\right\rangle=e^{-\I E_0t/\hbar}
          \frac{e^{\half\delta U(t)}
        \left|\Psi_0\right\rangle}{[\langle\Psi_0|e^{\half\delta U^\dagger(t)}
        e^{\half\delta U(t)}|\Psi_0\rangle]^{1/2}}\ ,
\end{equation}
where $\ket{\Psi_0}$ is the ground state, and $\delta U(t)$ is an
excitation operator that is written, for the case of bosons, in
exactly the same manner as the ground state wave function:
\begin{equation}
        \delta U(t)
        =\sum_i\delta u_1(\rvec_i;t)
        +\sum_{i<j}
          \delta u_2(\rvec_i,\rvec_j;t) + \ldots
\label{eq:deltaU}
\end{equation}

The amplitudes $\delta u_i(\rvec_1,\ldots\rvec_i;t)$ are determined
by the time-dependent generalization of the Ritz' variational principle:
\begin{equation}
  \frac{\delta}{\delta u_i(\rvec_1,\dots\rvec_i;t)}
  \int\!\! dt\,\langle\,\Psi(t)|H-\I\hbar\partial_t|\Psi(t)\rangle = 0\,.
\label{eq:eom}
\end{equation}

Assuming that $\delta U(t)$ is a small perturbation of the ground
state correlations, we can linearize the equations of motion for
$\delta u_i(\rvec_i,\dots;t)$, leading to the density--density
response function $\chi(k,\omega)$ from which we obtain the dynamic
structure function $S(k,\omega)=\Im m\,\chi(k,\omega)$:
\begin{equation}
\chi(q,\omega) =
        \frac{S(q)}{\hbar \omega - \varepsilon_{\rm F}(q) - \Sigma(q,\hbar\omega)}
        +
        \frac{S(q)}{-\hbar \omega -\varepsilon_{\rm F}(q) - \Sigma(q,-\hbar\omega)}\ ,
\label{eq:chi}
\end{equation}
where $\varepsilon_{\rm F}(k) = \hbar^2 k^2/2m S(k)$ is the Feynman excitation
spectrum, and the self-energy is given by an integral equation
\begin{equation}
        \Sigma(q,\hbar\omega) =
        \frac{1}{2}
        \int \frac{d^3k_1d^3k_2}{(2\pi)^3\rho}\,
        \frac{\delta({\bf q}-{\bf k}_1-{\bf k}_2)
        \left|\tilde V_3({\bf q};{\bf k}_1,{\bf k}_2)\right|^2}
        {\hbar\omega-\varepsilon_{\rm F}(k_1)
        -\Sigma(k_1,\hbar\omega-\varepsilon_{\rm F}(k_2))
        -\varepsilon_{\rm F}(k_2)-\Sigma(k_2,\hbar\omega-\varepsilon_{\rm F}(k_1))}\,.
\label{eq:sigma}
\end{equation}
$\tilde V_3(\kvec;\pvec,\qvec)$ is the three-phonon vertex
\begin{equation}
       \tilde V_3({\bf q};{\bf k}_1,{\bf k}_2)
       =\frac{\hbar^2}{2m}\sqrt{\frac{S(k_1)S(k_2)}{S(q)}}
        \left[
        {\bf q}\cdot {\bf k}_1\tilde X(k_1)+{\bf q}\cdot {\bf k_2}\tilde X(k_2)
        - q^2\tilde X_3({\bf q},{\bf k}_1,{\bf k}_2)\right]\,,
\label{eq:V3}
\end{equation}
where $\tilde X(k) = 1-1/S(k)$. $\tilde X_3({\bf q},{\bf k}_1,{\bf
  k}_2)$ is the fully irreducible three-phonon coupling matrix
element. In the simplest approximation, $\tilde X_3({\bf q},{\bf
  k}_1,{\bf k}_2)$ is replaced by the three--body correlation $\tilde
u_3(\qvec,\kvec_1,\kvec_2)$; this approximation ensures that
long--wavelength properties of the excitation spectrum are preserved
\cite{Chuckphonon}.  Improved calculations \cite{eomI} sum a 3-point
integral equation to ensure that exact properties of $\tilde X_3({\bf
  q},{\bf k}_1,{\bf k}_2)$ as $q \rightarrow 0$ {\em and\/} of the
Fourier transform $X_3({\bf r}_1,{\bf r}_2,{\bf r}_3)$ for $|{\bf
  r}_1-{\bf r}_2| \rightarrow 0$ and $|{\bf r}_1-{\bf r}_3|
\rightarrow 0$ are satisfied \cite{eomI}. We include these corrections
routinely, they have a visible effect only for wave vectors between
the maxon and the roton. The CBF-Brillouin-Wigner (CBF-BW)
approximation \cite{JacksonSkw} is obtained by omitting the
self-energy corrections in the energy denominator of
Eq. (\ref{eq:sigma}).

The implementation of the method outlined only briefly here has led to
an unprecedented agreement between theoretical predictions \cite{eomIII}
and experimental results \cite{skw4lett} which are still being
explored \cite{skwpress}.

\begin{figure}[H]
\centerline{\includegraphics[width=1.2\textwidth,angle=-90]{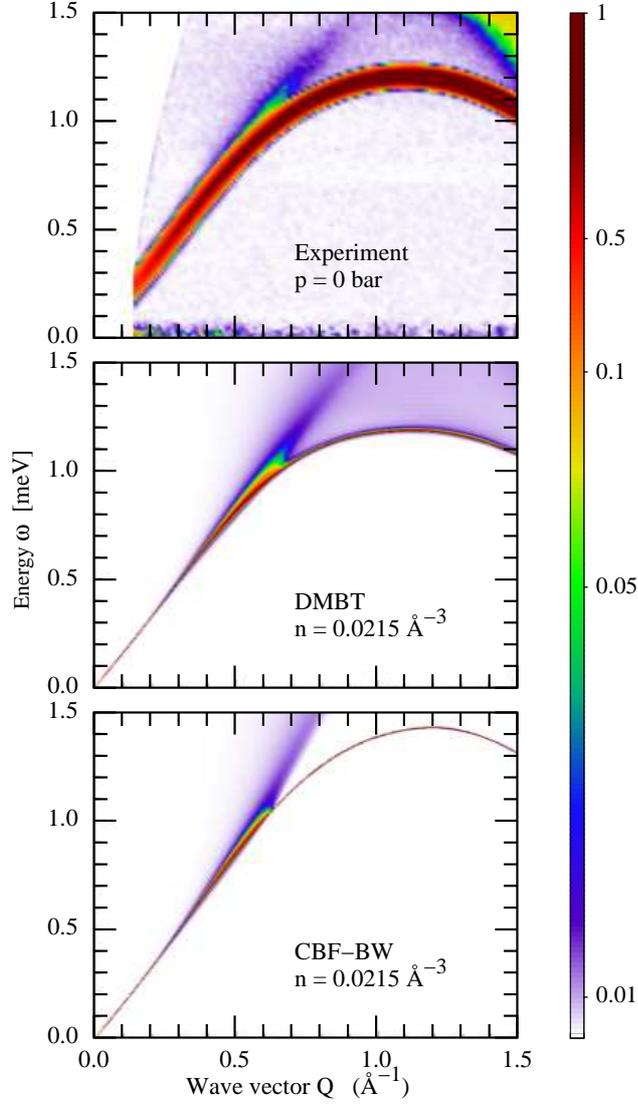}}
  \caption{A comparison of theoretical and experimental data for the
    dynamic structure function $S(k,\omega)$ from experiments
    \cite{skw4lett,skwpress} and two versions of our theory: The
    middle figure is based on the solution of the integral equation
    (\ref{eq:sigma}), the lowest figure shows the CBF-BW
    approximation; the only difference to previous work
    \cite{JacksonSkw} are more accurate input functions $S(k)$ and
    $X_3(\rvec_1,\rvec_2,\rvec_3)$.}
  \label{fig:skwfigure}
\end{figure}

In Fig. \ref{fig:skwfigure}, we compare the experimental dynamic
structure function $S(k,\omega)$ (top panel) \cite{skw4lett} with
results from DMBT calculations (middle panel) and CBF-BW calculation
\cite{JacksonSkw} (bottom panel). In the CBF-BW calculation we have,
of course, used the best available values for $S(k)$ in the
three-phonon vertex (\ref{eq:V3}) and included the irreducible part
$\tilde X_3(\qvec;\kvec_1,\kvec_2)$ as described in
Ref. \citenum{eomI}. Thus, the numerical values are different from
those of Ref. \citenum{JacksonSkw} but the physics described is
basically the same.

The most significant difference between the CBF-BW approximation and
the full solution of the integral equation (\ref{eq:sigma}) is the
void regime above the phonon-maxon-roton dispersion relation. This is
caused by the fact that, in the CBF-BW approximation, excitations
decay into Feynman phonons $\varepsilon_{\rm F}(k)$ and not into the
physical phonons. Furthermore, CBF-BW overestimates the excitation
energies in the maxon region, while DMBT agrees with the experiment.

\section{Phonon dispersion}
\label{sec:dispersion}

The theoretical phonon dispersion relation is obtained by solving the
implicit equation
\begin{equation}
  \varepsilon_0(k) = \varepsilon_{\rm F}(k) + \Sigma(k,\varepsilon_0(k))\,.
  \label{eq:disp}
\end{equation}
At long wave lengths, the phonon dispersion relation is
\begin{equation}
   \varepsilon_0(k) \approx \hbar c k (1-\gamma k^2)\label{eq:dispersion}
\end{equation}
where $c$ is the speed of sound and $\gamma$ is the {\em phonon
  dispersion coefficient\/}. If $\gamma<0$, we speak of anomalous
dispersion, long--wavelength phonons are damped.

To obtain the dispersion coefficient $\gamma$ we have fitted both experimental
and theoretical data by a polynomial of the form
\begin{equation}
  \varepsilon_0(k) = \hbar c k\left(1 - 
  \gamma k^2 + \alpha_3 k^3 + \alpha_4 k^4\right),
\label{eq:dispfit}
\end{equation}
from which the phonon dispersion coefficient was extracted.
The polynomial form turned out to be more flexible than the Pad\'e approximation
\cite{MarisRMP,PhysRevA.8.1980}
\begin{equation}
  \varepsilon_0(k) = \hbar c k\left(1 -
  \gamma k^2\frac{1-k^2/Q_a^2}{1+k^2/Q_b^2}\right)\,.
\label{eq:dispfitMaris}
\end{equation}
In particular, the Pad\'e approximation (\ref{eq:dispfitMaris}) does not
contain the term proportional to $k^3$ which can be calculated
analytically from the asymptotic form of the microscopic two-body
interaction. Assuming the typical asymptotic form $V(r) = C_6 r^{-6}$,
one arrives at \cite{Pitaevskii1970,Feenberg1971}
\begin{equation}
  \alpha_3 = \frac{\pi^2}{24}\frac{\rho}{mc^2}C_6\,.
  \label{eq:alpha3}
\end{equation}
We should note that fitting procedure should not be understood as a
rigorous low-$k$ expansion in the sense of a Taylor expansion of the
dispersion relation $\epsilon_0(k)$ around $k=0$, but rather as a fit
to the data in the theoretically and experimentally relevant
regime. The fit works well for $k < 0.6\,$\AA$^{-1}$.

To explain the fact that the above fitting procedure should not be
considered to be a rigorous Taylor expansion, we must review a little
more of the theoretical background. The Feynman spectrum
$\varepsilon_{\rm F}(k)$ is derived from a Bogoliubov formula
\begin{equation}
  \varepsilon_{\rm F}(k) = \sqrt{ t^2(k) + 2 t(k) \tilde V_{\rm p-h}(k)}
  \label{eq:efeyn}
\end{equation}
where $t(k) = \hbar^2 k^2/2m$ is the kinetic energy, and
\begin{equation}
  \tilde V_{\rm p-h}(k) \equiv \rho\int d^3r V_{\rm p-h}(r)e^{\I\kvec\cdot\rvec}
\end{equation}
is the ``particle-hole'' interaction or, in the language of Aldrich
and Pines \cite{Aldrich} the ``pseudopotential''. For what follows we
only need the property that, for large distances, $V_{\rm p-h}(r)$
falls off like the bare interaction, {\em i.e.\/} $V_{\rm p-h}(r)\sim
C_6 r^{-6}$ for $r\rightarrow\infty$.  Then, $\tilde V_{\rm p-h}(k)$
has the expansion
\begin{equation}
  \tilde V_{\rm p-h}(k) = \tilde V_{\rm p-h}(0+) + V_2 k^2 + \frac{\pi^2}{12}\rho C_6 k^3
\label{eq:vexpand}
\end{equation}
where $\tilde V_{\rm p-h}(0+) = \rho\int d^3r V(r) = mc^2$ is related
to the speed of sound, and $V_2$ is the second moment $V_2 =
-\frac{\rho}{6}\int d^3r V_{\rm p-h}(r) r^2$. Higher moments do not
exist due to the van der Waals tail.

\begin{figure}
\centerline{\includegraphics[width=0.5\textwidth,angle=-90]{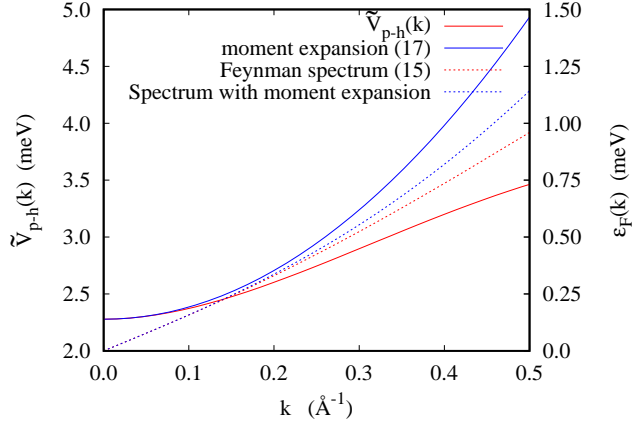}}
\label{fig:potfit}
\caption{The figure shows the full particle--hole interaction $\tilde
  V_{\rm p-h}(k)$ (solid red line, left scale) and the low-momentum
  expansion (\ref{eq:vexpand}) (solid blue line) at a density
  $\rho=0.0215$\AA$^{-3}$. Also shown are the Feynman dispersion
  relations obtained from the Bogoliubov relation (\ref{eq:efeyn}) for
  these two cases (red and blue dashed lines, right scale).}
\end{figure}

Fig. \ref{fig:potfit} shows the full $\tilde V_{\rm p-h}(k)$ the way
it is used in the Bogoliubov formula (\ref{eq:efeyn}) and the
expansion (\ref{eq:vexpand}) as calculated from the potential
moments. Evidently, the agreement is very good only in the regime
$0\le k\le 0.1\,$\AA$^{-1}$ which is inaccessible to neutron
scattering measurements.  The figure also shows the Feynman spectrum
$\varepsilon_{\rm F}(k)$ as obtained from the full $\tilde V_{\rm
  p-h}(k)$ and from the moment expansion. 

Both the DMBT and the CBF-BW results for $\gamma$ agree quite well
with the experimental data, whereas the Feynman approximation predicts
anomalous dispersion at all densities.

\begin{figure}
\centerline{\includegraphics[width=0.6\textwidth,angle=-90]{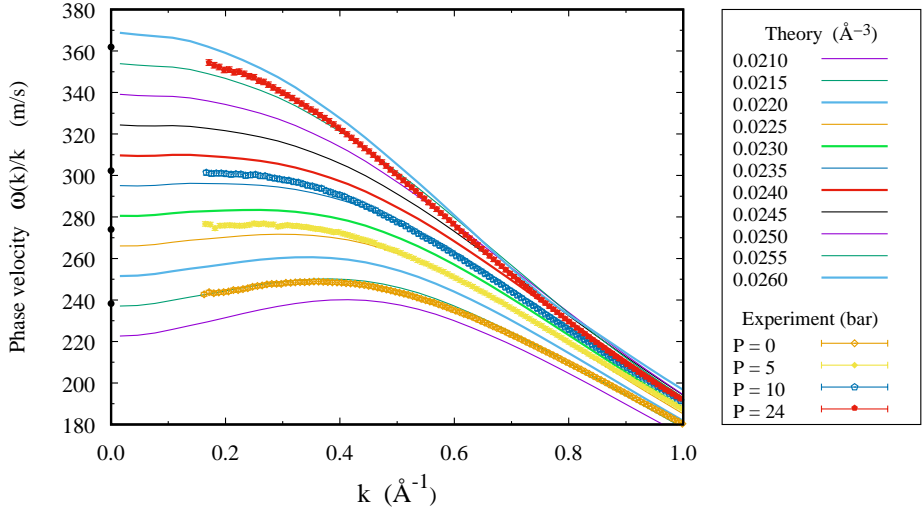}}
\label{fig:velocities}
\caption{Dispersion relation for several $^4$He densities, comparing
  experiment (markers for pressures P=0, 5, 10, 24 bar, corresponding
  to densities $0.0218, 0.0230, 0.0239$ and $0.0258$\AA$^{-3}$), and
  DMBT (lines, for several densities from $0.0210$ to
  $0.0260$\AA$^{-3}$ as indicated in the legend). Thick lines indicate
  theoretical curves for densities close to the experimental values
  given above. The black dots at $k=0$ indicate the sound velocities
  determined by ultrasonic techniques \cite{Abraham}.  }
\end{figure}

The calculated phase velocities $c = \varepsilon_0(k)/ \hbar k$ are 
given in Fig. \ref{fig:velocities} for densities covering the full 
range between the saturated vapor pressure and solidification. 
Also shown are experimental data   \cite{skw4lett,skwpress} 
at four pressures corresponding to the same the same density range. 
Clearly the agreement is excellent. Only a very small shift 
in density is observed between the theoretical calculations 
and the experimental results, already discussed in previous 
publications   \cite{skw4lett,skwpress}.

\begin{figure}
\centerline{\includegraphics[width=0.65\textwidth,angle=-90]{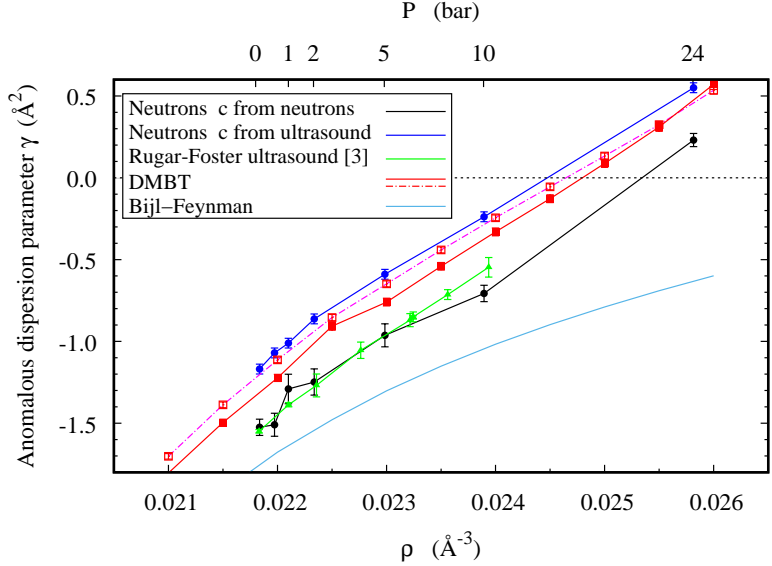}}
\caption{Dispersion coefficient $\gamma$ for several $^4$He densities,
  comparing neutron scattering experiments (blue and black dots and
  lines correspond to different data analyses described in the text),
  ultrasonic measurements \cite{Rugar1984} (green triangles) and
  DMBT results (red lines and solid and open squares). Similar results are
  found for the different fit ranges $0.0-0.6\,$\AA$^{-1}$
  (dash-dotted red line) and $0.2-0.6\,$\AA$^{-1}$ (solid red
  line). We also show the dispersion coefficient coming from the
  Feynman spectrum (\ref{eq:efeyn}) (light blue
  line). \label{fig:anom}}
\end{figure}

The polynomial expression (\ref{eq:dispfit}) provides an excellent fit 
of the theoretical curves in the small wave vector range. In practice, 
fits of the experimental curves were done in the range 
$0.18 < k < \ 0.6\,$\AA$^{-1}$, the lower bound being determined by the 
neutron detectors smallest angle. The highest bound was determined 
as the maximum wave vector where stable fits could be obtained using the 
form (\ref{eq:dispfit}). This was also verified for the theoretical curves;
for the latter, the fit range could be extended to $k=0$, 
without affecting significantly the results of the fits, as seen in  
Fig. \ref{fig:anom}. 

The experimental results have been analyzed using the polynomial
expression (\ref{eq:dispfit}) to determine the dispersion coefficient
$\gamma$ for several $^4$He densities.  The speed of sound $c$
determined by the fit agrees well with the well know values measured
by ultrasonic or thermodynamic techniques \cite{Abraham,CaupinJordi},
as can be seen in the extrapolations to $k=0$ in
Fig. \ref{fig:anom}. The agreement is not perfect, however, and this
affects the value of the next term in the expansion, the dispersion
coefficient $\gamma$. We thus show in Fig. \ref{fig:anom} the curves
for $\gamma$ determined using either the fitted values of $c$, or
the ultrasonic ones.

For completeness, we have also extracted other data from both the
calculations and the experiments. Table~\ref{tab:dispk} gives the
calculated coefficients $\hbar c$, the speed of sound $c$, the
Gr\"uneisen-Constant $u$ and the dispersion coefficient
$\gamma$. Static quantities like $c$ and $u$ can also be obtained from
Monte Carlo simulations,

\begin{table}[ht]
  \centerline{
    \begin{tabular}{|r | r r r r r | r r r |}
  \hline
    $\rho$ (\AA$^{-3}$)  & $\hbar c$\ (meV\AA)& $c$ (m/sec) & $u$ &
    $\gamma_{\rm F}$ & $\gamma$\ \  \ \ & $c$ &
    \ u &\ $\gamma$\ \ \ \\
    \hline
    &\multicolumn{4}{c}{JF-EL}&\multicolumn{1}{c|}{DMBT}&\multicolumn{3}{c|}{expt./DMC}\\
    \hline
    0.0210\ &\ 1.466 &\ 222.7 &\ 2.738 &\ -2.221 &\ -1.804\ \ &\ 211.5\ &\ 3.080 &\ -1.85\ \ \\
    0.0215\ &\ 1.561 &\ 237.2 &\ 2.628 &\ -1.946 &\ -1.495\ \ &\ 227.0\ &\ 2.944 &\ -1.56\ \ \\
    0.0220\ &\ 1.657 &\ 251.7 &\ 2.534 &\ -1.710 &\ -1.223\ \ &\ 242.6\ &\ 2.832 &\ -1.29\ \ \\
    0.0225\ &\ 1.752 &\ 266.2 &\ 2.453 &\ -1.510 &\ -0.906\ \ &\ 258.2\ &\ 2.738 &\ -1.04\ \ \\
    0.0230\ &\ 1.848 &\ 280.7 &\ 2.384 &\ -1.333 &\ -0.760\ \ &\ 274.0\ &\ 2.657 &\ -0.81\ \ \\
    0.0235\ &\ 1.944 &\ 295.3 &\ 2.323 &\ -1.182 &\ -0.541\ \ &\ 289.9\ &\ 2.588 &\ -0.60\ \  \\
    0.0240\ &\ 2.040 &\ 309.9 &\ 2.269 &\ -1.046 &\ -0.330\ \ &\ 305.9\ &\ 2.529 &\ \ \  \\
    0.0245\ &\ 2.136 &\ 324.6 &\ 2.222 &\ -0.925 &\ -0.127\ \ &\ 322.1\ &\ 2.476 &\ \ \  \\
    0.0250\ &\ 2.234 &\ 339.4 &\ 2.179 &\ -0.815 &\ 0.087\ \ &\ 338.5\ &\ 2.430 &\ \ \  \\
    0.0255\ &\ 2.331 &\ 354.2 &\ 2.141 &\ -0.713 &\ 0.311\ \ &\ 355.0\ &\ 2.388 &\ \ \ \\
    0.0260\ &\ 2.429 &\ 369.1 &\ 2.107 &\ -0.619 &\ 0.571\ \ &\ 371.7\ &\ 2.532 &\ \ \  \\
    \hline
    \end{tabular}}
  \caption{The table shows in columns 2-6 the calculated values $\hbar
    c$, the speed of sound $c$, the Gr\"uneisen constant $u$ and the
    phonon dispersion coefficient $\gamma_F$ in Feynman approximation,
    and as obtained by fitting the dispersion relation by the form
    (\ref{eq:dispfit}).  Col. 7-9 give the corresponding quantities
    obtained, as far as possible, from experiments or from Monte Carlo
    simulations. The experimental results for the dispersion
    coefficient $\gamma$ have been obtained by interpolating the data
    of Ref. \citenum{Rugar1984} at the densities in col. 1 by a
    quadratic polynomial.
\label{tab:dispk}}
\end{table}

Evidently, the agreement between the DMBT results and the experiment
is excellent at all densities. The dispersion coefficient $\gamma$
turns positive above $\rho\approx 0.0245$\AA$^{-3}$, meaning that
long--wavelength phonons can propagate freely only at high pressures.
The Feynman approximation predicts, on the other hand, a negative
dispersion coefficient at all densites.

\section{Phonon Mean Free Path}
\label{sec:MeanFreePath}

If $\gamma < 0$, a phonon of energy/momentum $(\hbar\omega, k)$ can
decay into two phonons of lower energy and longer wave length.  As a
consequence, the self--energy becomes complex on the phonon dispersion
relation. The onset of the imaginary part of an excitation of energy
and momentum $(\hbar\omega,k)$ is at the critical energy
$\hbar\omega_{\rm crit} = 2\varepsilon_0(k/2)$, where
$\varepsilon_0(k)$ is the dispersion relation calculated with DMBT,
see eq.~(\ref{eq:disp}).  At a corresponding critical wave number
$k_{\rm crit}$ determined by solving
$2\varepsilon_0(k/2)=\varepsilon_0(k)$ for $k$, a phonon can decay
into two phonons, parallel to the original phonon but with half the
wave number.  At this point the imaginary part of the self--energy is
\cite{bosegas,eomIII}
\begin{equation}
        \Sigma(k,\hbar\omega) =
        - \frac{|\tilde V_3({\bf k};-{\bf k}/2,-{\bf k}/2)|^2k}
{16\pi\rho\varepsilon_0'(k/2) }
        \sqrt{\frac{2\varepsilon_0(k/2)-\hbar\omega}
        {\varepsilon_0''(k/2)}}\,.
\label{eq:SigmaCrit3D}
\end{equation}
 
Phonons with lower wave numbers, $k<k_{\rm crit}$, have more decay
channels because the wave vectors of the produced phonons need not be
parallel to the original phonon; this angular spread is discussed
further below.  Phonons with wave numbers $k>k_{\rm crit}$ do not
decay into pairs of phonons and have infinite life-time in the DMBT
approximation.  However, higher-order processes, {\i.e.\/} decay into
three and more phonons, lead to a long, but finite life-time also for
$k>k_{\rm crit}$.

The life time of phonons with $k<k_{\rm crit}$ can be readily calculated from the
imaginary part of the self-energy,
\begin{equation}
  \tau(k) = \hbar\, \Im m[\Sigma(k,\hbar\varepsilon_0(k),k)]^{-1}
\label{eq:tau}
\end{equation}
The mean free path of a phonon, important for the suggested application of
superfluid $^4$He for dark matter detection mentioned above, can be
obtained from
\begin{equation}
  d(k) = v_g(k)\, \tau(k)
  \label{eq:d}
\end{equation}
where $v_g={d\omega(k)\over dk}$ is the group velocity.

Evidently, two things are needed for a reliable theoretical prediction
of the phonon life time $\tau(k)$ and of the mean free path $d(k)$.  (1)
An accurate dispersion coefficient is required for the low-momentum
{\em kinematics},
\begin{eqnarray}
\kvec&=&\qvec+\pvec\label{eq:momentum}\\
\varepsilon_0(k) &=&\varepsilon_0(q)+\varepsilon_0(p)\label{eq:energy}
\end{eqnarray}
which in particular determines the range of momenta where phonon
damping occurs.  (2) The three-phonon vertex $V_3$ is required for an
accurate damping {\em strength}.  The comparison of the dynamic
structure function between experiment and the DMBT result in
Fig.~\ref{fig:skwfigure} shows that DMBT is accurate enough that
eqns.(\ref{eq:tau}) and (\ref{eq:d}) indeed provide a reliable
theoretical prediction of phonon life time and mean free path.

\begin{figure}
  \centerline{\includegraphics[height=0.41\textwidth]{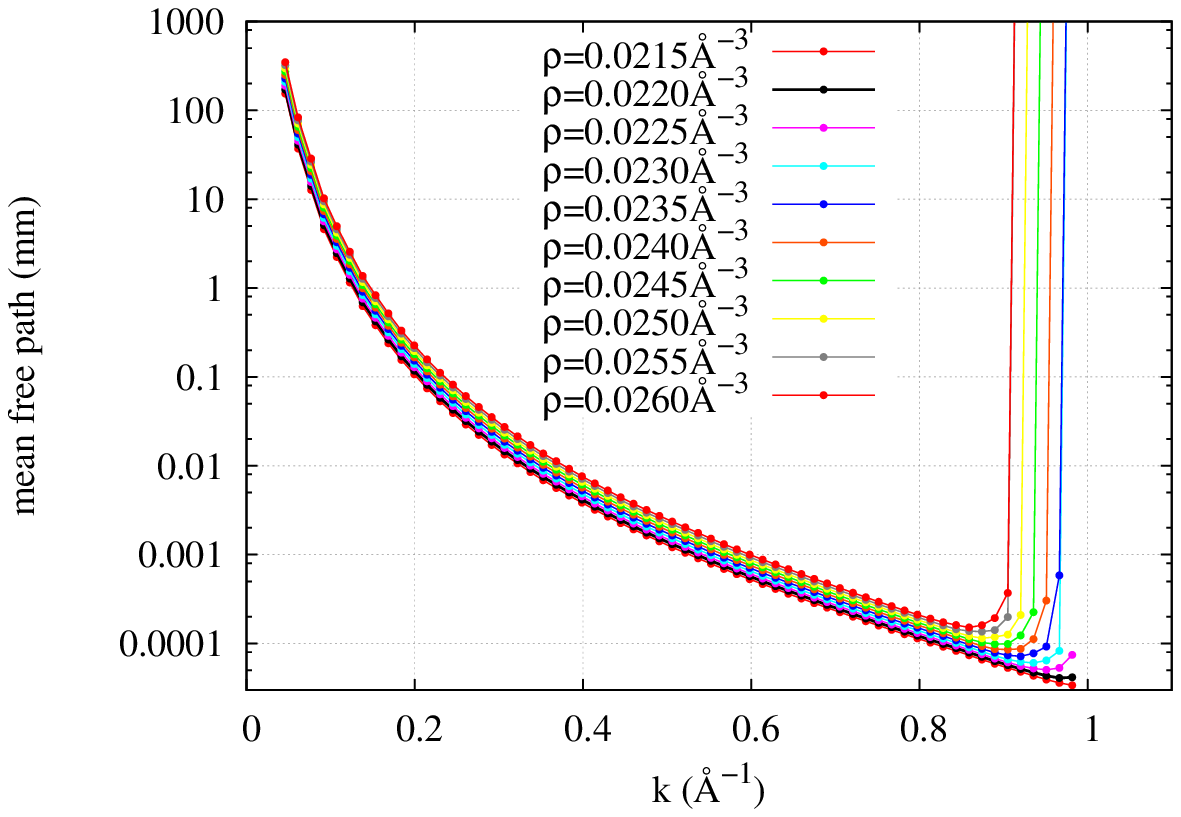}
  \includegraphics[height=0.41\textwidth]{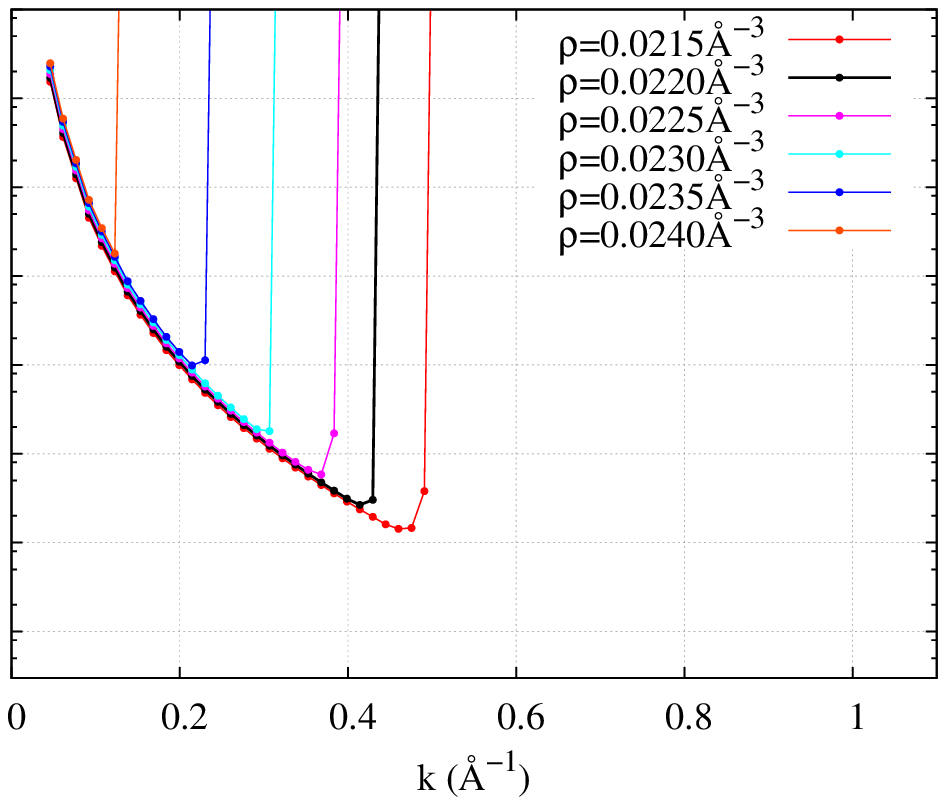}}
  \caption{The phonon mean free path $d(k)$ as function of the wave number
  	for different pressures.
  	The left panel shows the result in CBF-BW approximation and the right one
    shows the improved result obtained with the DMBT method which shows that
    the phonon mean free path depends strongly on the pressure.  \label{fig:decay}
}
\end{figure}

Figs. \ref{fig:decay} show our results for the phonon
mean free path $d(k)$ in both CBF-BW approximation
(left panel) and in DMBT (right panel) for several helium densities $\rho$.
The density range is smaller in the latter case, because the
dispersion relation is not anomalous anymore for
$\rho\gtrsim 0.0245$\AA$^{-3}$,
see the dispersion coefficient $\gamma$ shown in
Fig.~\ref{fig:anom}.  In the CBF-BW calculation phonons
decay by producing Feynman-phonons, which overestimates
the anomalous dispersion and which have a negative $\gamma$
for all densities considered here (see Fig.\ref{fig:anom}).
Therefore CBF-BW yields a finite mean free path even for
$\rho=0.0260$\AA$^{-3}$.  In fact $d(k)$ is almost independent
of $\rho$ and thus of the pressure in the CBF-BW approximation, while the much improved
DMBT result shows that the range of momenta
where phonons can decay strongly depends on $\rho$, hence on the pressure.
While CBF-BW predicts a minimal decay length of about $0.1\mu$m for
all densities, the DMBT results show that
the minimal decay length is $1-10\mu$m around the equilibrium density,
and increases significantly with density and pressure, because
$\gamma$ crosses zero and becomes positive at high density, Fig.\ref{fig:anom},
until the phonons do not decay anymore
for densities higher than $0.0240$\AA$^{-3}$,

While the DMBT calculation predicts a much smaller range of
wave numbers at which phonons are dampened, the values for
mean free path $d(k)$ at different densities are almost identical
for a given $k$, if there is damping at all.  In fact, the $d(k)$ values
predicted by CBF-BW and DMBT are essentially the same.
The reason is that the lifetime and thus the mean free path are
mostly determined by the vertex $V_3$, which is the same for CBF-BW and DMBT.
The strength of damping does depend on the first derivative of
the dispersion which, however, for the small $k$ range relevant here is
very similar in all approximations (Feynman, CBF-BW, and DMBT)
-- only the higher order derivatives, essential for the correct
decay kinematics, are sensitive to the approximation.

\section{Inelastic currents}

Anomalous dispersion is a prerequisite for phonon decay, but the
deviation of the dispersion relation from a linear dispersion is, in
the wave number regime under consideration here, hardly visible, see
Fig. \ref{fig:velocities}.  Therefore, a phonon decays into a
pair of more or less collinear phonons. The angular distribution of
these decay phonons can be obtained from the transport current, which
is the {\em second order\/} expectation value of the current operator
in the fluctuations $\delta U(\rvec_1,\ldots,\rvec_N,t)$
\begin{eqnarray}
  {\bf j}^{(2)}({\bf r},t) &\equiv&
  \frac{1}{4}\Bra{\Psi_0}\delta U^*\,\hat{\bf j}({\bf r})\,\delta U\Ket{\Psi_0}
\nonumber\\
  &=&\frac{\hbar N}{8m\I} \int d^3r_2 \ldots d^3r_N\Psi_0^2
  \left[\delta U^*(t)\nabla_1\delta U(t)-{\rm c.c.}\right]
\label{eq:SecondOrderCurrent}
\end{eqnarray}
The derivation of workable formulas is somewhat tedious, some
essential steps will be presented in the Appendix. A detailed
derivation for the case of impurity scattering off the $^4$He surface
may be found in Ref. \citenum{hscatt}, and for the case of impurity
and $^4$He scattering in Ref. \citenum{dropdyn}.

The decay of a phonon with wave vector $\kvec$ produces two phonons
according to the kinematics (\ref{eq:momentum}) and (\ref{eq:energy}).
We are interested in a measure for the
probability that a decay product is ejected
in direction $\evec$ ($\evec$ is a unit vector with Euler angles
$\theta$ and $\varphi$).  For this purpose we calculate
the rate of inelastic phonon current in direction $\evec$, for
which we obtain
\begin{equation}
\frac{d}{dt} j^{(2)}(\evec) =
  \frac{1}{mS(k)}\int \frac{d^3 q}{(2\pi)^3\rho}\, q\, \delta(\evec-{\qvec\over q})\,
  {\Im m}\,\sigma(\kvec,\qvec,\omega)
\label{eq:j2direction}
\end{equation}
The $\delta(\evec-{\qvec/ q})$ obviously selects only phonons with
wave vector $\qvec$ parallel to the direction of interest.  Here
$\sigma(\kvec,\qvec,\omega)$ is
\begin{equation}
  \sigma(\kvec,\qvec,\omega) =
  \frac{1}{2} \frac{\left|\tilde V(\kvec,\kvec-\qvec,\qvec)\right|^2}
  {\hbar\omega-\varepsilon_0(|\kvec-\qvec|) -\varepsilon_0(q) +\I\eta}
\end{equation}
The integration over $\qvec$ yields the self energy $\int
\frac{d^3q}{(2\pi)^3\rho} \sigma(\qvec;\kvec,\omega) =
\Sigma(\kvec,\omega)$, see Eq.~(\ref{eq:sigma}). For the calculation
of $\sigma(\kvec,\qvec,\omega)$, we have replaced the non--local
self--energy appearing in Eq. (\ref{eq:sigma}) by the dispersion
relation (\ref{eq:disp}), this is legitimate in the low momentum
regime under consideration.

\begin{figure}
  \centerline{\includegraphics[width=0.80\textwidth]{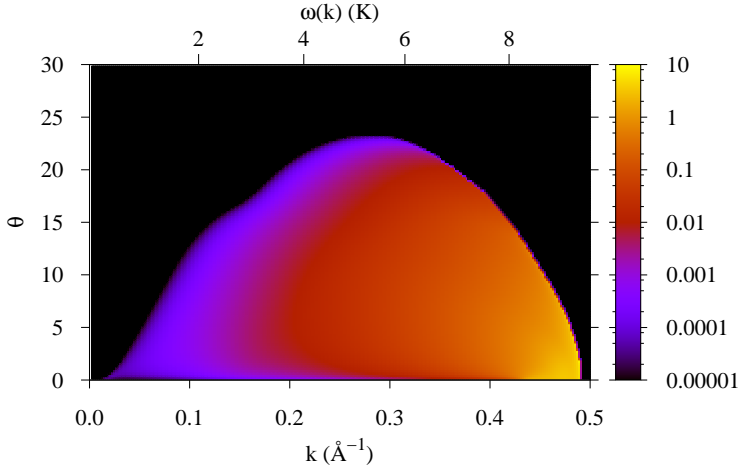}}
  \label{fig:theta}
  \caption{The angular dependence of the phonons produced by the decay
    of an initial phonon with wave number $k$. $\theta$ is the angle of the decay
    phonons with respect to the direction of the initial phonon.
    The density is $\rho = 0.0215$\AA$^{-3}$.}
\end{figure}

Fig. \ref{fig:theta} shows, at the density $\rho = 0.0215$\AA$^{-3}$,
the DMBT prediction for the rate with which a phonon with wave number
$\kvec$ decays into phonons with direction $\theta$ with respect to
$\kvec$ (obviously, the rate does not depend on $\varphi$).  The
figure summarizes both the kinematics and the strength of phonon
damping by decay into lower momentum phonons.  In agreement with the
decay length in the right panel of Fig. \ref{fig:decay}, the decay
rate is highest for phonons with momentum $k$ close to 0.5\AA$^{-1}$.
Fig.\ref{fig:theta} shows that these higher energy phonons decay into
lower energy phonons which lie in a narrow cone (small
$\theta$). Phonons with $k=0.2 - 0.3$\AA$^{-1}$ decay at a smaller
rate (have a larger decay length, see Fig.\ref{fig:decay}), but
generate phonons in a larger cone up to angles of $\theta=20^\circ$.

\section{Conclusion}

The dispersion and damping of low momemtum phonons in superfluid
helium-4 is of practical interest for cosmological applications in
weakly interacting particle detectors to test hypotheses for dark
matter.  We presented a comparison of the low momentum dispersion in
$^4$He between experimental data and the results from dynamic
many-body theory (DMBT) and find excellent agreement. The dispersion
obtained with DMBT is a significant improvement over the CBF
prediction in the full momentum and energy range of interest. This
allows us to make quantitative predictions for the decay of phonons
due to the anomalous dispersion without phenomenological or
experimental input parameters.  The phonon mean free path is similar
to CBF-BW results, but the CBF-BW approximation fails to predict the
pressure dependence, leading to decay lengths as short as 0.1$\mu$m.
This shortcoming is remedied by the DMBT which shows that the minimal
decay length is strongly pressure dependent.

\appendix
\section*{Transport current}

We present in this appendix the essential steps of the derivation of
the transport currents; the details of the derivation are rather
similar to the ones for the impurity currents derived in
Refs. \citenum{hscatt} and \citenum{dropdyn}. We derive the inelastic
current at the level of pair fluctuations, {\em i.e.\/} we truncate
the expansion (\ref{eq:deltaU}) at the two-body level. To derive the
integral equation (\ref{eq:sigma}) one needs to include fluctuations
to all orders \cite{eomIII}; our result will be plausible enough such
that we can avoid these complications.

As in the derivation of the linearized equations of motion
for the correlation operator (\ref{eq:deltaU}), leading to the
density--density response function
$\chi(k,\omega)$, Eq.~(\ref{eq:chi}), it is
convenient to apply an weak harmonic perturbation (e.g. induced by a neutron beam)
\begin{equation}
  U_{\rm ext}(\rvec;t) = \left[U_{\rm ext}(\rvec) e^{-\I\omega t} + U_{\rm ext}^*(\rvec) e^{\I\omega t}\right] e^{\eta t}
\end{equation}
As usual, the perturbation is switched on adiabatically with an infinitesimal rate $\eta>0$.
After linearization, the positive and negative frequency terms leads to a corresponding response of the
correlation operator $\delta U(t)$ split into (complex) positive and negative frequency terms; similarly
for the response of one-body, two-body densities, $\delta\rho_n$.

Inserting the correlation operator  (\ref{eq:deltaU}) at the two-body level
in the transport currents (\ref{eq:SecondOrderCurrent}) gives (we omit the
time dependence for clarity)
\begin{equation}
{\bf j}^{(2)}(\rvec) = \frac {\hbar}{4m\I}\left[
  \delta\rho_1^*(\rvec_1)\nabla_1\delta u_1(\rvec_1)
  +\int d^3r_2 \delta\rho_2^*(\rvec_1,\rvec_2)\nabla_1\delta u_2(\rvec_1,\rvec_2) \right]\,.
\label{eq:j2}
\end{equation}
where we have used that we are only interested in the time-averaged
currents, since only those are observable by a detector.

While looking very simple, the expression (\ref{eq:j2}) for the
current contains both the fluctuations of the correlation and of the
densities. In accordance with the derivation of the self energy
$\Sigma(k,\hbar\omega)$, summarized in the text, we express $\delta
u_1$ and $\delta\rho_2$ in terms of $\delta\rho_1$ and $\delta u_2$.
It turns out advantageous to introduce $\delta v_1(\rvec;t)$ such that
\begin{equation}
  \delta\rho_1(\rvec_1) = \rho\left[S*\delta v_1\right](\rvec_1)\,.
\end{equation}
where we define the convolution
product with the static structure function
\begin{equation}
  \left[S*f\right](\rvec_1) = f(\rvec_1) + \rho\int d^3r_2h(\rvec_1,\rvec_2)f(\rvec_2)\,.
\end{equation}
In linear order, we can then write
\begin{eqnarray}
\delta v_1(\rvec) &=& \delta u_1(\rvec) + \delta w_1(\rvec)\nonumber\\
\delta w_1(\rvec) &=& \rho \int d^3r_2 h(\rvec_1,\rvec_2)\delta
u_2(\rvec_1,\rvec_2)
\nonumber\\
&&+ \frac{\rho^2}{2}\int d^3r_2 d^3r_3
 Y(\rvec_1;\rvec_2,\rvec_3)\delta u_2(
\rvec_2,\rvec_3)\,,
\label{eq:dvdef}
\end{eqnarray}
where the $Y(\rvec_1;\rvec_2,\rvec_3)$ is the set of those contributions
to the three-body distribution function that is non-nodal in the point
$\rvec_1$; details can be found in Ref.~\cite{eomIII}.

In all further manipulations, we use the ``convolution approximation''
which is the simplest approximation for high-order distribution
functions that maintains the exact long-wavelength properties.  In
convolution approximation, $Y(\rvec_1;\rvec_2,\rvec_3)$ is simply
\begin{equation}
  Y(\rvec_1;\rvec_2,\rvec_3) = h(\rvec_1,\rvec_2) h(\rvec_1,\rvec_3)\,.
  \label{eq:Yconv}
\end{equation}

Using Eqs.~(\ref{eq:dvdef}) and (\ref{eq:Yconv}) and expressing
$\delta\rho_2(\rvec_1,\rvec_2)$ in terms of $\delta\rho_1(\rvec)$ and
the fluctuation of the pair distribution function $\delta
g(\rvec_1,\rvec_2)$,
\[
  \delta\rho_2(\rvec_1,\rvec_2) = 
  \rho\left[\delta\rho_1(\rvec_1)+\rho_1(\rvec_2\right]\,g(\rvec_1,\rvec_2)
+ \rho_2 g(\rvec_1,\rvec_2)
\]
we obtain
\begin{eqnarray}
  {\bf j}^{(2)}(\rvec;t) &=& \frac {\hbar}{4m\I}
  \Biggl\{\delta\rho^*_1(\rvec_1)\biggl[\nabla_1\delta v_1(\rvec_1)-\rho
      \int d^3r_2\delta u_2(\rvec_1,\rvec_2) \nabla_1 h(\rvec_1,\rvec_2)\nonumber\\
      &&\qquad-\frac{\rho^2}{2}\nabla_1 \int d^3r_2 d^3r_3
      h(\rvec_1,\rvec_2)h(\rvec_1,\rvec_3)
      \delta u_2(\rvec_2,\rvec_3)      \biggr]\nonumber\\
  &&\qquad+\rho\int d^3r_2 h(\rvec_1,\rvec_2)\nabla_1\delta u_2(\rvec_1,\rvec_2)
  \delta\rho_1^*(\rvec_2)\nonumber\\
  &&\qquad+\rho^2\int d^3r_2 \delta g^*(\rvec_1,\rvec_2)\nabla_1\delta u_2(\rvec_1,\rvec_2)\Biggr\}\,.
\end{eqnarray}

Finally we eliminate $\delta g(\rvec_1,\rvec_2)$ using the convolution
approximation:
\begin{equation}
  \delta g(\rvec_1,\rvec_2) = \int d^3r_3 h(\rvec_1,\rvec_3)h(\rvec_2,\rvec_3)
  \delta\rho_1(\rvec_3)
+ \left[S*\delta u_2*S\right](\rvec_1,\rvec_2)\,.
\end{equation}
which yields the length expression for the current
\begin{eqnarray}
{\bf j}^{(2)}(\rvec_1)
&=&
  \frac {\hbar}{4m\I}
  \delta \rho^*_1(\rvec_1)\nabla_1\delta v_1(\rvec_1) \label{eq:jfirst}\\
&+&
  \frac {\hbar\rho}{4m\I}
  \int d^3r_2
  \nabla_1\left[\delta u_2*S\right](\rvec_1,\rvec_2)h(\rvec_1,\rvec_2)
  \delta\rho_1^*(\rvec_2)  \label{eq:jsecond}\\
&+&
  \frac {\hbar\rho}{4m\I}\delta \rho^*_1(\rvec_1)
  \int d^3r_2 \left[S*\delta u_2\right](\rvec_1,\rvec_2)\nabla_1 h(\rvec_1,\rvec_2)  \label{eq:jthird}\\
&+&
  \frac {\hbar}{4m\I}
  \int d^3r_2\left[S*\delta u_2^* *S\right](\rvec_1,\rvec_2)
  \nabla_1\delta u_2(\rvec_1,\rvec_2) \label{eq:jfourth}
\end{eqnarray}
The first term (\ref{eq:jfirst}) is the current in Feynman
approximation which is the elastic channel. The second and third
terms, (\ref{eq:jsecond}) and (\ref{eq:jthird}), are correlations
between the density fluctuation $\delta\rho_1$ and the pair
correlation fluctuation $\delta u_2$.  We will show below that only
the fourth term (\ref{eq:jfourth}) contains the decay rate of an
elementary excitation (in the above derivation given in Feynman
approximation).

We now need explicit expressions for the one- and two-body
fluctuations. These come from solving the equations of motion,
formally given by Eq. (\ref{eq:eom}), in detail given in
Refs.~\citenum{hscatt,eomIII}. We can assume that the perturbation
$U_{\rm ext}(\rvec)$ is a plane wave with wave vector ${\bf q}$ which
induces a plane wave density fluctuation, propagating with phase
velocity given by the dispersion relation, $\omega_0(q) =
\varepsilon_0(q)/\hbar$:
\begin{equation}
  \delta\rho(\rvec;t) = \lambda\rho e^{\I(\qvec\cdot\rvec-\omega_0(q) t)} e^{\eta t}
\label{eq:deltarho1}
\end{equation}
where $\lambda\ll 1$ is an arbitrary strength factor.

The two-body correlation fluctuations are then coupled to the density
fluctuation according to the equations of motion.  Their spatial
Fourier transform is given by
\begin{eqnarray}
  \tilde u_2(\kvec_1,\kvec_2) &=&
  \frac{\lambda(2\pi)^3\delta({\bf q}+{\bf k}_1+{\bf k}_2)}
       {\varepsilon_0(q)-\varepsilon_0(k_1)-\varepsilon_0(k_1) +\I\hbar\eta}
  \frac{\tilde V({\bf q};{\bf k}_1,{\bf k}_2)}
       {\sqrt{S(q)S(k_1)S(k_2)}}e^{-\I\omega_0(q) t) +\eta t}
\label{eq:x2ofu1}
\end{eqnarray}
We finally use Eq. (\ref{eq:deltarho1}) for $\delta\rho_1(\rvec)$ and
Eq. (\ref{eq:x2ofu1}) for $\delta u_2(\rvec_1,\rvec_2)$ in the current
${\bf j}^{(2)}(\rvec)$ above.  The fourth term (\ref{eq:jfourth}) which
becomes
\[
  \lambda^2\frac{\hbar}{4m}\int \frac{d^3 k_1d^3k_2}{(2\pi)^3\rho}
  \delta(\qvec+\kvec_1+\kvec_2)
    \,\kvec_1\, \frac{1}{S(q)}
    \frac{\left|\tilde V(\qvec;\kvec_1,\kvec_2)\right|^2}
       {(\varepsilon_0(q) - \varepsilon_0(k_1)-\varepsilon_0(k_2))^2 + \hbar^2\eta^2}\, e^{2\eta t}
\]
In the limit of adiabatically switching on the perturbation, $\eta\to 0$, we
need to calculate the {\em rate} with which the inelastic current is
generated, {\em i.e.\/} the time derivative.
Using
\begin{equation}
  \frac{d}{dt}\frac{e^{2\eta t}}
       {(\varepsilon_0(q) - \varepsilon_0(k_1)-\varepsilon_0(k_2))^2+
         \hbar^2\eta^2}\rightarrow 
	\frac{2\pi}{\hbar}\,\delta(\varepsilon_0(q) - \varepsilon_0(k_1)-\varepsilon_0(k_2))
\end{equation}
we obtain the rate of the total inelastic current $\frac{d}{dt}{\bf
  j}^{(2)}$ apart from the arbitrary strength factor $\lambda^2$.
Selecting only the inelastic phonon current (the decay products)
in a specific direction ${\bf e}$
relative to the incoming phonon with wave vector ${\bf q}$, we obtain the rate
Eq.~(\ref{eq:j2direction}).  Note that the two mixed term
contributions to ${\bf j}^{(2)}(\rvec)$, Eqs. (\ref{eq:jsecond}) and
(\ref{eq:jthird}), do not have a finite contribution to the rate in
the adiabatic limit.

\bibliography{papers}
\bibliographystyle{spphys}

\end{document}